\documentstyle [12pt,dina4,epsf]{article}
\begin{document}
\parindent=0cm
\parskip=1.5mm

\def\bi{\begin{list}{$\bullet$}{\parsep=0.5\baselineskip
\topsep=\parsep \itemsep=0pt}}
\def\ei{\end{list}}

\def\phi{\varphi}
\def\-{{\bf --}}
\def\vm{v_{max}}
\def\ta{\tilde a}
\def\ts{\tilde s}
\def\tr{\tilde r}
\newcommand{\s}{\sigma}
\newcommand{\la}{\lambda}
\newcommand{\eps}{\varepsilon}
\newcommand{\al}{\alpha}
\newcommand{\nn}{{\cal N}}

\newcommand{\be}{\begin{equation}}
\newcommand{\ee}{\end{equation}}
\newcommand{\bea}{\begin{eqnarray}}
\newcommand{\eea}{\end{eqnarray}}

\newcommand{\nonu}{\nonumber\\}

\renewcommand{\thefootnote}{\fnsymbol{footnote}}

\parindent=0cm
\begin{center}
  {\LARGE\bf Car-oriented mean-field theory}
\end{center}
\begin{center}
  {\LARGE\bf for traffic flow models}
\end{center}
\vskip1.8cm \renewcommand{\thefootnote}{\fnsymbol{footnote}}
\setcounter{footnote}{1}
\begin{center}
  {\Large Andreas Schadschneider$^{1}$ and Michael Schreckenberg$^{2}$}
\end{center}
\vskip1.3cm
\begin{center}
  $^{1}$ Institut f\"ur Theoretische Physik\\ Universit\"at zu K\"oln\\ 
  D--50937 K\"oln, Germany\\
  email: {\tt as@thp.uni-koeln.de}
\end{center}
\begin{center}
  $^{2}$ Mathematik/FB 11\\ 
  Gerhard-Mercator-Universit\"at Duisburg\\ 
  D--47048 Duisburg, Germany\\
  email: {\tt schreck@traf3.math.uni-duisburg.de} 
\end{center}
\begin{center}
\today
\end{center}

\vskip1.3cm \vskip2cm {\large \bf Abstract}\\[0.2cm]
We present a new analytical description of the cellular automaton
model for single-lane traffic. In contrast to previous approaches
we do not use the occupation number of sites as dynamical variable
but rather the distance between consecutive cars. Therefore certain 
longer-ranged correlations are taken into account and even a mean-field
approach yields non-trivial results. In fact for the model with
$\vm=1$ the exact solution is reproduced. For $\vm=2$ the fundamental
diagram shows a good agreement with results from simulations.
\hspace{.4cm} 

\vskip2cm
\vfill
\pagebreak
Despite a large number of publications about the cellular automaton
approach to traffic flow (see e.g.\ \cite{Jue} and references therein) 
only a few of those deal with a systematic
analytical description. Most works make use of large-scale computer
simulations which can be carried out very efficiently for this class
of models. Nevertheless, analytical results -- exact or approximate --
may give important information relevant for a complete understanding
of those models.

The most important exact result is certainly the solution
of the model for $\vm =1$ \cite{ss}. This result has been obtained
using $n$-cluster-approximation \cite{ss,ssni}, i.e.\ an improved mean-field 
theory taking into account correlations between $n$ neighbouring sites.
For $\vm =1$ already the 2-cluster approximation is exact \cite{ss}.
For higher velocities the $n$-cluster-approximation for small $n$ already
yields very good results for the so-called fundamental diagram (flow-density
relationship) \cite{ss,ssni}.

In computer simulation studies there are in principle two different 
approaches \cite{ssni} called site-oriented and car-oriented\footnote{In
\cite{ssni} this approach has been called particle-oriented.}.
In the site-oriented approach the state of the system is specified by
storing the state of each cell which can either be empty or occupied
by a single car with velocity $v=0,1,\ldots,\vm$.
In the car-oriented approach on the other hand one stores the velocity 
of each car and the distance to the next car ahead.

Since in the cluster-approximation corresponds to a site-oriented approach
this analogy inspired us to investigate an analytical description based
on the car-oriented approach, the so-called car-oriented mean-field theory
(COMF)\footnote{A brief account of some preliminary results has already 
been given in \cite{ssjuel}.}. The COMF already takes into account some 
longer-ranged correlations so that one can hope that it yields at least
a good approximation.

For completeness we briefly repeat the definition of the CA model for
single-lane traffic flow \cite{ns} in the following. The street is divided 
into $L$ cells of a certain length (for realistic applications 7.5 meters) 
which can be occupied by at most one car or be empty. The cars have an 
internal parameter ('velocity') which can take on only integer values 
$v=0,1,2,\ldots,\vm$. The dynamics of the model are described by the
following update rules for the velocities and the motion of 
cars \cite{ns}. In the first step all cars with velocities $v_i < \vm$
are accelerated by one velocity unit, $v_i\rightarrow v'_i=v_i+1$. 
The following step describes the slowing down due to other cars and 
prevents accidents. All cars with velocities $v_i' > d_i$ (where $d_i$
is the number of free cells in front of car $i$) decelerate to
velocity $v''_i=d_i$. The last step in the velocity update is a
randomization effect taking into accout several aspects of the 
driver's behaviour, e.g.\ fluctuations in driving style, overreaction
at breaking, and retarded acceleration: Every car with velocity $v''_i>0$ will
slow down one unit with probability $p$, i.e.\ $v''_i \stackrel{p}{\to} 
v'''_i = v''_i-1$. In the final step the car then moves $v'''_i$ sites.
These four rules, referred to as step 1 to step 4 in the following, 
are applied to all cars at the same time (parallel dynamics). 
\vskip0.5cm
{\underline{COMF for $v_{max} =1$}}

We denote the probability to find at time $t$ (exactly) $n$ empty sites 
in front of a vehicle by $P_n(t)$. As in \cite{ss,ssni} we change the
order of the update steps to 2-3-4-1. This change has to be taken into
account when calculating the flux $f(c,p)$. It has the advantage that
after step 1 there are no cars with velocity 0, i.e.\ all cars have 
velocity 1. 
The time evolution of the probabilities $P_n(t)$ can conveniently be
expressed through the probability $g(t)$ ($\bar g(t)=1-g(t)$) that
a car moves (does not move) in the next timestep. 

In order to find the time evolution of the $P_n(t)$ we first determine
from which configurations at time $t$ a given state at time $t+1$ 
could have been evolved under the rules 2-3-4-1. Take for instance a car 
-- called second car in the following -- which has $n>1$ free sites in front,
i.e.\ its distance to the next car ahead (called first car in the following)
is $n+1$ sites. This configuration might have evolved from four different
configurations at time $t$, depending on whether $(i)$ both cars moved
in the timestep $t\to t+1$ (which happens with probability $qg(t)$), 
$(ii)$ both cars did not move (with probability $p\bar g(t)$), $(iii)$
only the first car moved (with probability $pg(t)$) or $(iv)$ only the 
second car moved (with probability $q\bar g(t)$). This means that the
second car at time $t$ had either $n$ free site in front (cases $(i)$ and
$(ii)$), or $n-1$ free sites (case $(iii)$), or $n+1$ free sites (case $(iv)$).

The special cases $n=0,1$ can be treated in a analogous fashion. 
In this way one obtains the time evolution of the probabilities as
\bea
P_0(t+1)&=& \bar g(t)\left[P_0(t)+qP_1(t)\right],\label{evol1a}\\
P_1(t+1)&=& g(t)P_0(t) + \left[qg(t)+p\bar g(t)\right]P_1(t) 
          + q\bar g(t)P_2(t),
   \label{evol1b}\\
P_n(t+1)&=& pg(t)P_{n-1}(t) + \left[qg(t)+p\bar g(t)\right]P_n(t)
          + q\bar g(t)P_{n+1}(t).
\label{evol1c}
\eea

A car will move in the next timestep if there is at least one empty
cell in front of it (probability $\sum_{n\geq 1} P_n(t)$) and if it
does not decelerate in the randomization step 3) (probability $q=1-p$).
Therefore the probabilities $g(t)$ and $\bar g(t)$ are given by
\bea
g(t)&=&q\sum_{n\geq 1} P_n(t) = q[1-P_0(t)]\nonu
\bar g(t)&=&P_0(t)+p\sum_{n\geq 1} P_n(t) = p+qP_0(t)
\label{gdef1}
\eea
where we have used the normalization
\be
\sum_{n\geq 0} P_n(t) = 1.
\label{norm1}
\ee

The probabilities can also be related to the density $c=N/L$ of cars. Since
each car which has the distance $n$ to the next one in front of him
'occupies' $n+1$ sites we have the following relation:
\be
\sum_{n\geq 0}(n+1)P_n(t) = \frac{1}{c}\ .
\label{dens1}
\ee

Here we are mainly interested in the stationary state ($t\to\infty$) with
$\lim_{t\to\infty} P_n(t)=P_n$. In order to determine the probabilities
in the stationary state we introduce the generating function
\be
P(z) = \sum_{n=0}^\infty P_nz^{n+1}.
\ee
After multiplying the corresponding equation in (\ref{evol1a})-(\ref{evol1c}) 
by $z^{n+1}$ and summing over all equations one finds an explicit expression 
for the generating function,
\be
P(z) 
=\frac{q(\bar g + gz)zP_0}{q\bar g-pgz} 
\label{genfct}.
\ee
The normalization condition (\ref{norm1}) and the density relation
(\ref{dens1}) imply that the generating function has to satisfy 
\be
P(1) = 1 \qquad\qquad {\rm and} \qquad \qquad 
P'(1) = \frac{1}{c},
\ee
where $P'(z)$ denotes the derivative of $P(z)$.

Using (\ref{genfct}) it is easy to obtain the probabilities explicitly:
\bea
  P_0&=&\frac{2qc-1+\sqrt{1-4qc(1-c)}}{2qc},\nonumber\\ 
  P_n&=&\frac{P_0}{p}\left(\frac{pg}{q\bar g}\right)^n = \frac{P_0}{p}
       \left(\frac{p(1-P_0)}{P_0+p(1-P_0)}\right)^n \qquad\qquad (n\geq 1)
  \label{Pnexpl}
\eea
where we already used (\ref{dens1}) to express $P_0$ through the density
$c$ of cars.

To obtain the fundamental diagram we have to calculate the flux.
It is given by $f(c,p)=cg=qc(1-P_0)$ from which one recovers the 
exact result \cite{ss,ssni}
\be
f(c,p)=\frac{1-\sqrt{1-4qc(1-c)}}{2}.
\label{fund1}
\ee

In \cite{ss,ssni} we expressed the exact solution in terms of the
pair probabilities $P(n_j,n_{j+1})$ to find two neighbouring sites
$j$ and $j+1$ in the state $(n_j,n_{j+1})$. Here $n_j=0$ denotes an
empty site and $n_j=1$ a site occupied by a car (with velocity 
1)\footnote{Cars with velocity 0 do not exist after the acceleration step.}.
In \cite{ss,ssni} it was shown that probabilities for larger
clusters factorize, i.e.\ $P(n_1,\ldots,n_L)=\prod_{j=1}^{L-1}P(n_j,n_{j+1})$.
The 2-cluster probabilities are related to the $P_n$ through
\bea
P(1,1)&=& cP_0 ,\nonumber\\
P^2(1,0)&=& c(1-c)P_1 ,\\
P(0,0)&=& (1-c)\frac{P_{n+1}}{P_n}\qquad (n\geq 1).\nonumber
\eea
The factors $c$ and $1-c$ appear due to the different normalization of
the $P_n$ and $P(n_j,n_{j+1})$. The $P_n$ are normalized by the number
of cars whereas the $P(n_j,n_{j+1})$ are normalized by the number of
sites.

The fact that the COMF yields the exact result comes not unexpected 
since the COMF takes into account all relevent correlations for the 
case $\vm=1$ where only nearest-neighbour correlations are non-trivial 
\cite{ss,ssni}. Although the model for $\vm >1$ is know to exhibit a 
different behaviour \cite{ssni} the COMF might yield good results in this case 
because it takes into longer-ranged correlations.
In the following we will investigate the case $\vm=2$.
\vskip0.5cm
{\underline{COMF for $v_{max} =2$}}

The case $\vm=2$ can be treated in a similar way as $\vm=1$. However, it 
is now necessary to introduce two different functions $P_n(t)$ and $B_n(t)$
describing the probabilities to find exactly $n$ empty sites in front of 
a car with velocity 1 and 2, respectively. Proceeding as in the case
$\vm=1$ we find the evolution equations. For the stationary state the 
probabilities obey the equations
\bea
P_0&=&g_0\left[P_0+B_0\right],\label{P0}\\
P_1&=&g_1\left[P_0+B_0\right]+pg_0\left[P_1+B_1\right],
\label{P1}\\
P_2&=&g_2\left[P_0+B_0\right]+pg_1\left[P_1+B_1\right]
+pg_0P_2,\label{P2}\\
P_3&=&pg_2\left[P_1+B_1\right]+pg_1P_2+pg_0P_3,
\label{P3}\\
P_n&=&pg_2P_{n-2}+pg_1P_{n-1}+pg_0P_n,\qquad\qquad (n\geq 4)
\label{Pn}
\eea
and
\bea
B_0&=&qg_0\left[P_1+B_1+B_2\right],\label{B0}\\
B_1&=&qg_1\left[P_1+B_1+B_2\right]+g_0\left[qP_2
+pB_2+qB_3\right],\\
B_2&=&qg_2\left[P_1+B_1+B_2\right]+g_1\left[qP_2
+pB_2+qB_3\right]
+g_0\left[qP_3+pB_3+qB_4\right],\label{B2}\\
B_n&=&g_2\left[qP_{n-1}+pB_{n-1}+qB_n\right]+g_1\left[
qP_n+pB_n+qB_{n+1}\right]\nonu
& &\quad +g_0\left[qP_{n+1}+pB_{n+1}
+qB_{n+2}\right].\qquad\qquad\qquad \qquad  (n\geq 3).\label{Bn}
\eea

The probabilities $g_\alpha$ that a car moves $\alpha$ sites ($\alpha 
=0,1,2$) in the next timestep are given by:
\bea
g_0&=&P_0+B_0+p\sum_{n\geq 1}P_n+pB_1,\nonu
g_1&=&q\sum_{n\geq 1}P_n+qB_1+p\sum_{n\geq 2}B_n,\label{gdef2}\\
g_2&=&q\sum_{n\geq 2}B_n.\nonumber
\eea
We just mention here that it is possible to derive the identities
$g_0=\sum_{n\geq 0}P_n$ and $g_1+g_2=\sum_{n\geq 0}B_n$ from (\ref{gdef2})
and (\ref{P0}-\ref{Bn}). Using the normalization
\be
\sum_{n\geq 0}\left[P_n+B_n\right] =1
\label{norm2}
\ee
we have $g_0+g_1+g_2=1$.

The conservation of density leads to the constraint
\be
\sum_{n\geq 0}(n+1)\left[P_n +B_n\right] = \frac{1}{c}\ .
\label{dens2}
\ee

We introduce the generating functions
\bea
P(z) &=& \sum_{n=0}^\infty P_n z^{n+1},\\
B(z) &=& \sum_{n=0}^\infty B_n z^{n+1}.
\eea
These functions have to satisfy $P(1)+B(1)=1$ and $P'(1)+B'(1)=\frac{1}{c}$
due to (\ref{norm2}) and (\ref{dens2}), respectively.

After multiplication with $z^{n+1}$ and summation over all the equations
(\ref{P0}-\ref{Bn}) one finds
\bea
P(z)&=&\frac{g(z)}{1-pg(z)} \left[ (qP_0 + B_0)z + pB_1z^2 \right],\\
B(z)&=&\frac{zg(z)}{z^2-(q+pz)g(z)} \left[ qP(z) -qB_0 -(qP_0+pB_0+qB_1)z  
 + (q-p)B_1z^2 \right],\nonu
\eea
where we have introduced the function $g(z) = g_0 + g_1z + g_2z^2$.
Note that this function satisfies $g(1)=1$ and $g'(1)=\frac{1}{c}f(c,p)$
is just the average velocity of the vehicles.

If one expresses the sums appearing in (\ref{gdef2}) by $P(z=1)$ and 
$B(z=1)$ one obtains the following relations
\bea
B_1 &=& \frac{q}{p}g_0-\frac{1}{p}(B_0+qP_0),\label{B1alt}\\
g_1 &=& p(1-P_0)+(q-\frac{p}{q})B_0+(1-\frac{p}{q})B_1,\\
g_2 &=& q(1-P_0)-(1+q)B_0-B_1.
\eea
With these relations also the normalization condition $P(1)+B(1)=1$ 
is satisfied. Now we can express the generating function completely in
terms of the two probabilities $P_0$ and $B_0$ only, since 
$g_0=\frac{P_0}{P_0+B_0}$ from (\ref{P0}). 

At this point it is surprising that we are still left with two unknowns 
$P_0$ and $B_0$ since we only have one free parameter, the density $c$.
In the following we determine a relation
between $P_0$ and $B_0$ from analytic properties of $B(z)$. Thus the
generating functions depend only on one free parameter, e.g.\ $P_0$, which
then can be related to the density via $P'(1)+B'(1)=\frac{1}{c}$. 

The denominator of $B(z)$ can be rewritten as $z^2-(q+pz)g(z) = 
pg_2(1-z)(z-s_+)(z-s_-)$ where $s_\pm$ are given by
$s_\pm = \frac{1}{a}\left[1\pm \sqrt{1+\frac{qg_0}{pg_2}a^2}\right]$ with
$a=\frac{2pg_2}{g_0+qg_1}$. Thus the denominator of $B(z)$ has three zeroes, 
$z=1$ and $z=s_\pm$. Two of these are located in the unit circle since 
$|s_-| \leq 1$. These zeroes have to be cancelled by corresponding zeroes of 
the numerator since $B(z)$ has to be analytic in the unit circle (otherwise one
would not have $\lim_{n\to\infty} B_n = 0$). It is easy to see that the
numerator indeed has a zero at $z=1$. Demanding that it
also has a zero at $z=s_-$ we find the missing relation between
$P_0$ and $B_0$,
\be
qP(s_-) -qB_0 - (qP_0+pB_0+qB_1)s_- + (q-p)B_1s_-^2 = 0.
\ee
Owing to (\ref{dens2}) we can regard the generating functions $P(z)$
and $B(z)$ as functions of the density $c$ only. The fundamental
diagram can then be obtained using the following expression for the  
flux:
\be
f(c,p)=c[g_1+2g_2].
\label{flux2}
\ee
Results are shown in Fig.\ 1 for $p=0.1$ and Fig.\ 2 for $p=1/2$. 
For $p=0.1$ we find an excellent agreement of the two curves.
For $p=0.5$ we still find an excellent agreement for small densities 
($c < 0.2$) and high densities ($c>0.5$). Only near the maximum there are 
deviations. For comparision we also
have included in Fig.\ 2 results from the site-oriented approach, i.e.\
the $n$-cluster approximation. The COMF result is much better than the
2-cluster result and comparable to the 3-cluster approximation. 
It seems that the COMF tends to overestimate the flux whereas
the $n$-cluster approximation yields a lower bound for the flux \cite{ssni}.
The result for $p=0.1$ shows, however, that the COMF result is not
systematically larger than the simulation results.

For small densities the average distance between the cars is large.
Therefore correlations between cars and neighbouring empty sites are 
much more important than those between two cars. These correlations are
better described by the COMF which is the reason why the COMF is
superior to the cluster approach for small cluster sizes in this regime.

The agreement between simulations and the COMF result is very good
up to densities close to the critical density $c_{crit} \approx 0.19$
for $p=1/2$ \cite{esss}. Here correlations between distances become important. 
It is to be expected that a combination of the cluster approach and 
the COMF will give a much better agreement of the calculated fundamental
diagram and the simulations in the region near the critical point and the
flux maximum.

We also have applied the COMF to generalizations of some modified models 
\cite{taka,benja,fukui}. The models discussed in \cite{taka,benja} have
different modified acceleration rules which for $\vm=1$ break the 
'particle-hole' symmetry, i.e.\ the fundamental diagrams are no 
longer symmetric with respect to $c=0.5$. It turns out that the COMF
gives are good agreement with simulations, but is no longer exact (even
for $\vm=1$). A full account of these results will be given elsewhere.

\noindent
{\bf Acknowledgements:} Part of this work has been performed within
the research program of the Sonderforschungsbereich 341 
(K\"oln-Aachen-J\"ulich). We like to thank N.\ Rajewsky and L.\ Santen
for helpful discussions.


\newpage
\begin{figure}
\vskip 130mm
\includegraphics{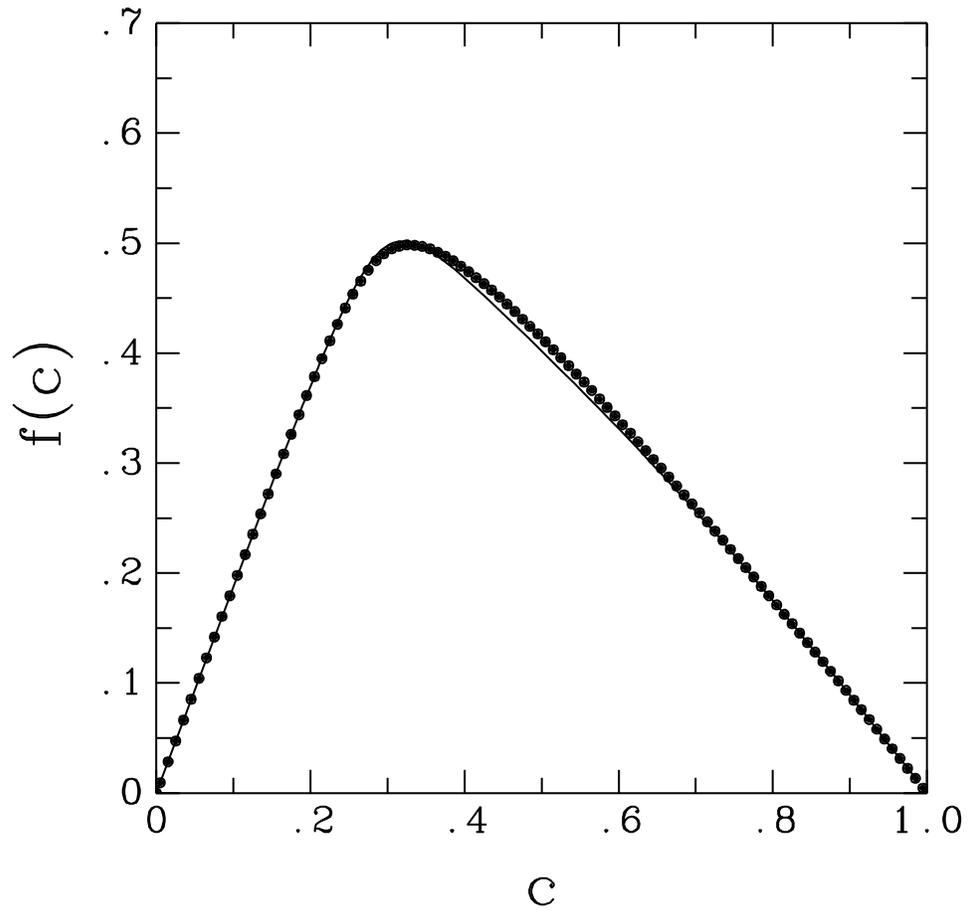}
\caption{Fundamental diagram for $\vm =2$ and $p=0.1$. The comparison of
  the COMF result (full line) with results from computer simulations 
($\bullet$) shows an excellent agreement.}
\label{fig1}
\end{figure}
\
\newpage
\begin{figure}
\vskip 60mm
\includegraphics{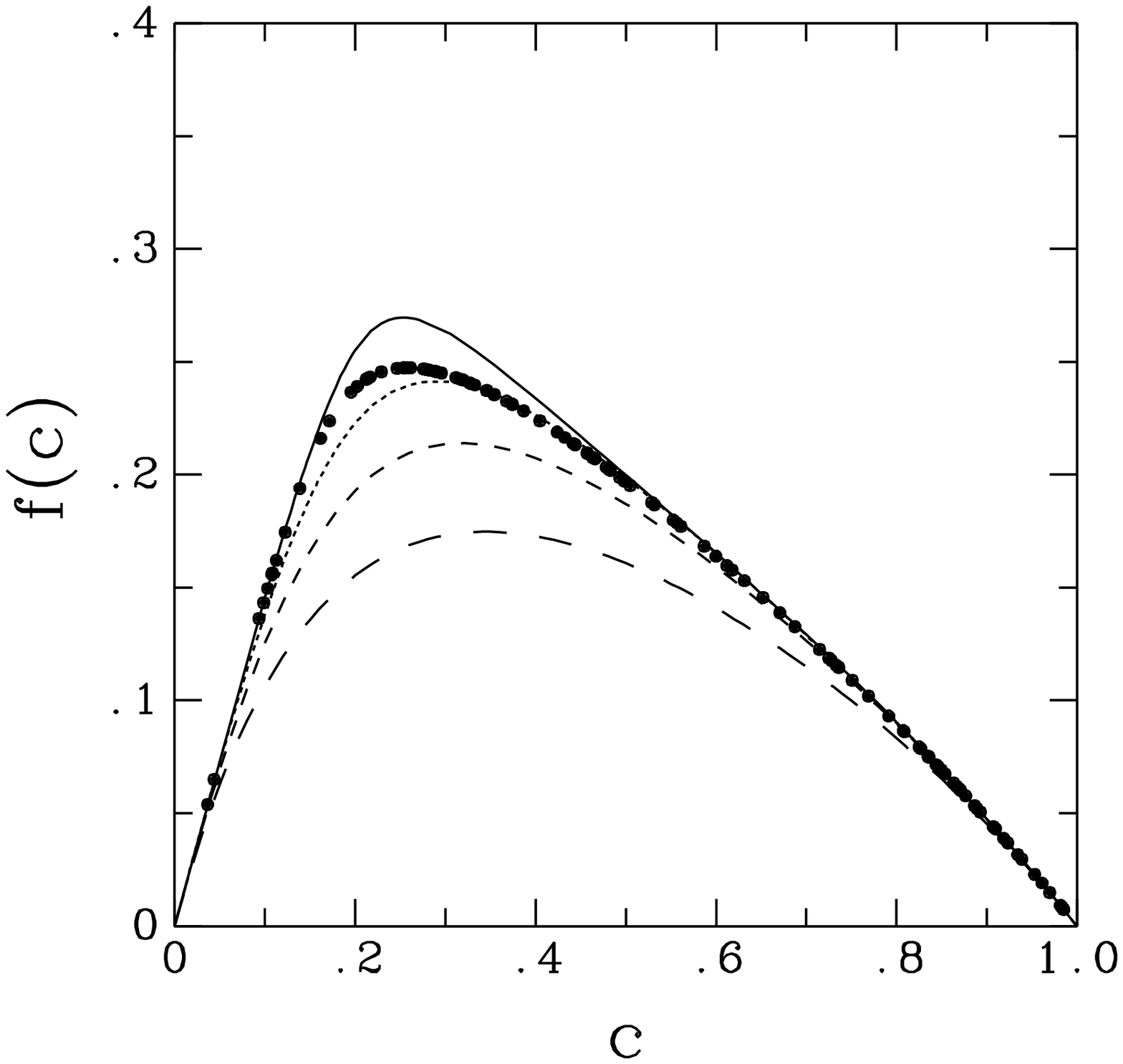}
\caption{Fundamental diagram for $\vm =2$ and $p=1/2$. The full
line is the COMF result. For comparison the results from computer 
simulations ($\bullet$) and the $n$-cluster approximation for 
$n=1,2,3$ (broken lines) are also shown.}
\label{fig2}
\end{figure}


\begin{thebibliography}{99}

\bibitem{Jue} D.E.\ Wolf, M.\ Schreckenberg, A.\ Bachem (Eds.): 
{\em Traffic and Granular Flow}, World Scientific, Singapore (1996)

\bibitem{ss} A.\ Schadschneider, M.\ Schreckenberg: J.\ Phys.\ {\bf A26}, 
L679 (1993)

\bibitem{ssni} M.\ Schreckenberg, A.\ Schadschneider, K.\ Nagel, N.\ Ito: 
Phys.\ Rev. {\bf E51}, 2939 (1995)

\bibitem{ssjuel} A.\ Schadschneider, M.\ Schreckenberg in \cite{Jue}

\bibitem{ns} K.\ Nagel, M.\ Schreckenberg: J.\ Phys.\ I France\ {\bf 2}, 2221
(1992)

\bibitem{esss} B.\ Eisenbl\"atter, L.\ Santen, A.\ Schadschneider, 
M.\ Schreckenberg: (to be published)

\bibitem{taka} M.~Takayasu, H.~Takayasu: Fractals {\bf 1}, 860 (1993)

\bibitem{benja} S.C.\ Benjamin, N.F.\ Johnson, P.M.\ Hui: J.\ Phys.\ 
{\bf A29}, 3119 (1996)

\bibitem{fukui} M.\ Fukui, Y.\ Ishibashi: J.\ Phys.\ Soc.\ Jpn.\ {\bf 65},
1868 (1996)

\end{thebibliography}
\end{document}